\documentclass[10pt,a4paper,prl,reprint,superscriptaddress]{revtex4-1}
\usepackage[T1]{fontenc} \usepackage{pstricks} \usepackage{subfigure}
\usepackage[normalem]{ulem}
\usepackage{graphicx}     
\usepackage{amsbsy}
\usepackage{amsfonts}
\usepackage{amsmath}
\usepackage{amssymb}
\usepackage{fixmath}
\usepackage[colorlinks]{hyperref}
\usepackage{physics}
\usepackage{color}

\hypersetup{
	bookmarksnumbered,
	pdfstartview={FitH},
	citecolor={darkgreen},
	linkcolor={blue},
	urlcolor={darkblue},
	pdfpagemode={UseOutlines}}
\definecolor{darkgreen}{RGB}{20,100,20}
\definecolor{darkblue}{RGB}{0,0,130}
\definecolor{darkred}{rgb}{.8,0,0}

\providecommand*{\I}{\mathrm{i}} 

\renewcommand{\vec}[1]{\mathbold{#1}}

\newcommand{\nab}{\mathbold{\nabla}}

\begin{document}
\title{Breathing mode of a Bose-Einstein condensate immersed in a Fermi sea}
\author{Piotr T. Grochowski}
\email{piotr@cft.edu.pl}
\affiliation{Center for Theoretical Physics, Polish Academy of Sciences, Aleja Lotnik\'ow 32/46, 02-668 Warsaw, Poland}
\author{Tomasz Karpiuk}
\email{t.karpiuk@uwb.edu.pl}
\affiliation{Wydzia{\l} Fizyki, Uniwersytet w Bia{\l}ymstoku,  ul. K. Cio{\l}kowskiego 1L, 15-245 Bia{\l}ystok, Poland}
\author{Miros{\l}aw Brewczyk}
\email{m.brewczyk@uwb.edu.pl}
\affiliation{Wydzia{\l} Fizyki, Uniwersytet w Bia{\l}ymstoku,  ul. K. Cio{\l}kowskiego 1L, 15-245 Bia{\l}ystok, Poland}
\author{Kazimierz Rz\k{a}{\.z}ewski}
\email{kazik@cft.edu.pl}
\affiliation{Center for Theoretical Physics, Polish Academy of Sciences, Aleja Lotnik\'ow 32/46, 02-668 Warsaw, Poland}

\begin{abstract}
By analyzing breathing mode of a Bose-Einstein condensate repulsively interacting with a polarized fermionic cloud, we further the understanding of a Bose-Fermi mixture recently realized by Lous et al. [\textit{Phys. Rev. Lett.} \textbf{120}, 243403].
We show that a hydrodynamic description of a domain wall between bosonic and fermionic atoms reproduces experimental data of Huang et al. [\textit{Phys. Rev. A} \textbf{99}, 041602(R)].
Two different types of interaction renormalization are explored, based on lowest order constrained variational and perturbation techniques.
In order to replicate nonmonotonic behavior of the oscillation frequency observed in the experiment, temperature effects have to be included.
We find that the frequency down-shift is caused by the fermion-induced compression and rethermalization of the bosonic species as the system is quenched into the strongly interacting regime.
\end{abstract}
\maketitle

\textit{Introduction}---Mixtures are routinely encountered in everyday life---from alloys and polymers to colloids and biological cell systems, various combinations of constituents have been studied to gain some advantageous physical properties.
It is no surprise that with the advent of the quantum era, novel multicomponent systems have entered the stage as an exciting alternative to thoroughly investigated classical structures.
Indeed, quantum mixtures have proved to be a fruitful playground for both theoretical and experimental physicists~\cite{Pethick2008,Pitaevskii2016}.

At the heart of multicomponent physics lies the analysis of the interaction between the mixture's constituents and how it affects overall properties of the system.
Specifically, strong repulsion between components may cause their spatial separation, heralding a phase transition and a change of order.
However, investigating subtle interplay between two strongly interacting constituents at the thin layer separating them is considerably more challenging than studying bulk properties of fully mixed compound, where an intercomponent overlap stays large and is easier to experimentally probe.

In a quantum science, for a long time, most of the interest was aimed at various phase-separated states of superfluid helium~\cite{Ebner1971} and solid-state settings.
However, introduction of experimental platform of ultracold gases opened a route towards increasingly better control and fine tuning of investigated systems~\cite{Anderson1995,Davis1995,Bradley1995}.
Widely utilized Feshbach resonances have allowed to freely adjust interaction between components in a quantum mixture, providing a highly clean environment for precise measurements of many-body and collective excitations~\cite{Chin2010}.

The latter are usually firstly investigated as an early probe of the properties of the multicomponent system~\cite{Mewes1996,Jin1996}.
Two types of collective modes are usually distinguished -- surface and compression ones.
An excitation is of surface type, when the volume of gas remains unchanged, in contrast to the other type.
Surface modes have been utilized to study e.g. collisionless to collisional crossover in both Bose~\cite{Stamper-Kurn1998,Buggle2005} and Fermi~\cite{Altmeyer2007,Wright2007} gases.
On the other hand, the response of the system under compression is usually used to study an equation of state of a given sample~\cite{Stringari1996,Chevy2002,Kinast2004,Bartenstein2004,Altmeyer2007a}.

It has been shown that intercomponent interaction can cause plethora of effects on collective oscillations of the system, yielding e.g. damping and frequency shifts~\cite{Busch1997,Esry1998,Ho1998,Hall1998a,Vichi1999,Bijlsma2000,Maddaloni2000,Capuzzi2001,Gensemer2001,Yip2001,Goral2002,Pu2002,Svidzinsky2003,Liu2003,Deconinck2004,Rodriguez2004,Navarro2009,Pixley2015,Shen2015,Wilson2016,Wu2018a}.
The particular role of phase separation in such problems has been extensively studied in Bose-Bose mixtures, in which description can be carried out accurately at the mean-field level~\cite{Papp2008a,Tojo2010,McCarron2011,Stamper-Kurn2013,Wacker2015,Wang2016,Lee2016}.
The introduction of fermionic species provides a greater challenge, as much stronger interaction is needed to overcome the mean-field value of the kinetic energy in the system~\cite{Molmer1998,Viverit2000,Roth2002,Zaccanti2006,Ospelkaus2006a,Lous2018,DeSalvo2019,Huang2019}.
The involvement of beyond-mean-field corrections cannot be therefore neglected, as shown in the case of repulsive two-component Fermi gas that undergoes phase separation~\cite{Stoner1933,Jo2009,Pilati2010,Massignan2011,Chang2011,Pekker2011,Sommer2011,Sanner2012,Massignan2014,Trappe2016,Valtolina2016,Grochowski2017a,Amico2018,Karpiuk2019}.

In this Letter, we study a binary mixture of Bose-Einstein condensate (BEC) of potassium-41 and polarized Fermi sea of lithium-6, that was recently realized and studied by the Innsbruck group~\cite{Lous2018,Huang2019}.
Both components are optically trapped and their interaction is tuned by the means of Feshbach resonances.
The boson-boson s-wave interaction, characterized by the scattering length $a_{bb}$, is kept constant, while interspecies interaction ($a_{bf}$) is finely tuned, allowing to access phase-separated regime.

The mixture is initially prepared as a very weakly interacting state ($a_{bb} \approx 60.9 a_0$ and $a_{bf} \approx 60 a_0$) in an elongated trap with the aspect ratio of $\lambda=7.6$.
The temperature is kept very low for fermions, staying at ca. $0.1 T_f$ for each investigated setting.
However, the condensed fraction of bosons is below $0.5$, suggesting the need of inclusion of the thermal cloud to the analysis.
After thermalization, to excite the breathing mode of the bosonic cloud, Bose-Fermi scattering length is alternately changed by the means of short radio-frequency pulses that switch the internal state of potassium atoms~\cite{Matthews1998,Pollack2010}.
The interaction is firstly quenched into value of $a_{bf} = 700 a_0$, then after half the radial breathing mode period of unaccompanied bosons, it is switched back to base weak interaction, $a_{bf} = 60 a_0$.
Such a procedure is repeated once more and then the gas evolves in the presence of ultimate interaction strength, $a_{bf}$.
Next, the frequency of a breathing mode is measured during couple of oscillation periods.

The theory presented in the original paper~\cite{Huang2019}, based on two approaches, adiabatic Fermi sea (AFS) and full phase separation (FPS) models, manages to qualitatively reproduce the up-shift of the frequency for weak-to-moderate interaction strengths.
However, it quantitatively overestimates the frequency and does not provide an explanation for a nonmonotonic behavior of the curve for a very strong repulsion.
By combining nonzero temperature classical fields description of cold Bose gas and hydrodynamics-derived pseudo-Schr\"{o}dinger representation of fermions, we acquire an approach that allows us to study dynamics of the Bose-Fermi mixture in a fully quantum way.
Our method provides a quantitative description of this previously unexplained behavior and suggests a way to further verify its predictions.

The Letter is organized as follows.
First, we present considerations about the energy spectrum of an uniform repulsive Bose-Fermi mixture within two different beyond-mean-field approaches -- lowest order constrained variational method (LOCV) and perturbative expansion by Viverit and Giorgini (VG).
Bose-Bose interaction is additionally refined by the inclusion of Lee-Huang-Yang (LHY) correction.
Then, coupled time equations are presented and nonzero temperature classical fields approximation is revisited.
With the methodology being established, we present the predictions for the breathing mode of such a mixture and explain what happens with it in a strong interaction regime.
The Letter is closed by the recapitulations and final remarks.

\textit{Energy of uniform mixture}---The starting point is the evaluation of the energy density of an uniform mixture of bosons and fermions with given single-particle densities, $n_b$ and $n_f$, respectively. 
The mean-field terms constitute of kinetic energy of fermions, $t_f^0=6^{5/3} \hbar^2 \pi^{4/3} n_f^{5/3}/20m_f $, Bose-Bose interaction energy, $\epsilon_{bb}^0=\frac{2 \pi \hbar^2}{m_b} a_{bb} n_b^2 $, and Bose-Fermi interaction energy, $\epsilon_{bf}^0=\frac{2 \pi \hbar^2}{\mu} a_{bf} n_b n_f $, where $m_{b(f)}$ is mass of the bosonic (fermionic) atom and $\mu=m_b m_f/(m_b+m_f)$ is the reduced mass.
The first refinement comes from a better treatment of the kinetic energy.
Two terms are added: kinetic energy of bosons, $t_b=\frac{\hbar^2}{2 m_b}(\nab \sqrt{n_b})^2$ and von Weizs\"acker correction to the kinetic energy of fermions, $\Delta t_f=\frac{\xi \hbar^2}{2 m_f}(\nab \sqrt{n_f})^2$ with $\xi=1/9$~\cite{Weizsacker1935,Kirzhnits1957}.
The latter is necessary to describe the fermionic density near a domain wall~\cite{Trappe2016}.
Then, $t_f=t_f^0+\Delta t_f$.
The first correction to the bosonic interaction energy comes in the form of the zero-point energy of the Bogolyubov vacuum of the bosonic system, the famous Lee-Huang Yang term, $\epsilon_{LHY}=C_{LHY} a_{bb}^{5/2} n_b^{5/2}$ with $C_{LHY} = 256 \sqrt{\pi} \hbar^2 / (15 m_b)$~\cite{Lee1957,SCHUTZHOLD2006,Lima2011}.

From this point, we use two different approaches for calculating beyond-mean-field corrections to the interaction energy of bosons and fermions.
The first method is based on the second-order perturbation theory, firstly considered in Ref.~\cite{Albus2002} and generalized by Viverit and Giorgini~\cite{Viverit2002}.
It yields the interaction energy, $\epsilon_{bf}^{VG} = C_{bf} a_{bf}^2 n_b n_f^{4/3} A(w,\alpha)$ with $C_{bf} = (6 \pi^2)^{2/3} \hbar^2 /2 m_f$ in terms of dimensionless parameters $w=m_b/m_f$ and $\alpha = 16 \pi n_b a_{bb} / (6 \pi^2)^{2/3} n_f^{2/3}$ .
The explicit form of the function $A(w,\alpha)$ is given in the Supplemental Materials (SM)~\cite{Rakshit2019}.

The other approach is not a perturbative one and in general can provide reliable results even in a strongly interacting regime.
Lowest order constrained variational (LOCV) method assumes an explicit symmetric term for the relative two-body correlations of s-wave interacting particles in the form of Jastrow factor, $\Psi \sim  \prod_{i,j} f(|\vec{r}_i-\vec{r}_j|)$~\cite{Pandharipande1973,Pandharipande1977,Cowell2002,Taylor2011,Yu2011}.
Such a trial function is also widely used in other variational approaches, mainly of quantum Monte Carlo type~\cite{Bertaina2013a}.
It was successfully applied in cold atoms systems, e.g. in repulsive Fermi-Fermi and attractive Bose-Fermi mixtures~\cite{Taylor2011,Yu2011}.

Within this approximation, the Bose-Fermi interaction energy can be then put in a simple form, $\epsilon_{bf}^{LOCV}= \frac{\hbar^2}{4 \mu} (6 \pi^2)^{2/3} n_f^{2/3} n_b B(\eta)$, where $B(\eta)$ is a numerically evaluated (see details in SM) function of dimensionless parameter $\eta = (k_f a_{bf})^{-1}$, where $k_f=(6 \pi^2 n_f)^{1/3}$.
The boson-boson interaction energy is also renormalized at the perturbative level, yielding $\epsilon_{bb}^{LOCV} = \frac{2 \pi \hbar^2}{m_b} a_{bb} n_b^2 (1+4 D(\eta)) $, where $D(\eta)$ can be numerically evaluated~\cite{Yu2011}.
The details of computations are presented in SM.

The comparison between two approaches allows us to check how important going beyond the mean field is in the considered Bose-Fermi mixture and how well different renormalizations of energy spectrum fare in dealing with the experimental results.
We use local density approximation (LDA) to describe the mixture in the harmonic trap, $V_b=\frac{1}{2} m_b \omega_b^2(\rho^2+\lambda^2 z^2)$ and $V_f=\frac{1}{2} m_f \omega_f^2 (\rho^2+\lambda^2 z^2)$, for bosons and fermions, respectively.
Then, the three energy spectra of a trapped Bose-Fermi mixture, mean field, LOCV, and VG read:
\begin{widetext}
\begin{align} \label{energy}
E_{MF} =& \int \dd^3 \vec{r} \left( t_f (\vec{r}) + t_b (\vec{r})  + \epsilon_{bb}^0 (\vec{r}) + \epsilon_{bf}^0 (\vec{r}) + n_b V_b (\vec{r}) + n_f V_f (\vec{r})  \right)  \nonumber \\
E_{LOCV}=& \int \dd^3 \vec{r} \left( t_f (\vec{r}) + t_b (\vec{r})  + \epsilon_{bb}^{LOCV} (\vec{r}) + \epsilon_{LHY} (\vec{r}) + \epsilon_{bf}^{LOCV} (\vec{r}) + n_b V_b (\vec{r}) + n_f V_f (\vec{r})   \right) \nonumber \\
E_{VG}=& \int \dd^3 \vec{r} \left( t_f (\vec{r}) + t_b (\vec{r})  + \epsilon_{bb}^{0} (\vec{r})+\epsilon_{LHY} (\vec{r}) + \epsilon_{bf}^0 (\vec{r})   + \epsilon_{bf}^{VG} (\vec{r}) + n_b V_b (\vec{r}) + n_f V_f (\vec{r}) \right)
\end{align}
\end{widetext}

\textit{Coupled time evolution equations}---We decide to represent bosons with the usual Bose field, however in the classical fields approximation (CFA)~\cite{Brewczyk2007,Blakie2008,Proukakis2008}.
In the CFA, the quantum Bose field $\widehat{\psi_b}(\vec{r},t) = \sum_{\vec{p}} \psi_{\vec{p}}^b (\vec{r}) \hat{a}_{\vec{p}} (t)$ describing the bosonic system is replaced by the classical Bose field $\sqrt{N_b} \psi_b(\vec{r},t) = \sum_{\vec{p}} \alpha_{\vec{p}} (t) \psi_{\vec{p}} (\vec{r},t). $
It constitutes an extension of the Bogolyubov approach, in which only $\vec{p}=0$ mode is macroscopically occupied.
This mode is then described by a single complex amplitude $\alpha_0$, that is governed by the Gross-Pitaevskii equation (GPE).
In the CFA, under the physically satisfied assumption of a finite system, each of the classical amplitudes $\alpha$ is macroscopically occupied and evolves under GPE.
The $|\alpha_0|^2$, the largest eigenvalue of the averaged single particle density matrix, is interpreted as a condensed fraction of bosons and higher amplitudes are part of the thermal cloud.
Such a description have been successfully employed for many problems in the past.

One of the ways to derive GPE is to start from the hydrodynamic description of Madelung~\cite{Madelung1927}.
Similar approach can also be readily utilized in the case of fermions, in the spirit of recent advances in the quantum continuum mechanics~\cite{Tao2009,Gao2010} or directly from quantum kinetic equations of motion~\cite{Rakshit2019a}.
The starting point is to represent fermions by a mean field pseudo-wave function, $\psi_f=\sqrt{n_f} \exp(\I\frac{m_f}{\hbar}\chi)$, where  ${\nab\chi=\vec v}$ gives irrotational velocity field of the collective motion~\cite{Trappe2016,Grochowski2017a,Gawryluk2017}.
By assuming that Hamiltonian is of the mean field form (\ref{energy}) and casting the time evolution equations in hydrodynamic (Euler-Lagrange) form, one can obtain coupled nonlinear pseudo-Schr\"odinger equations by the means of inverse Madelung transformation:
\begin{align}\label{evo}
i \hbar \partial_t \psi_j = &\Big[ -\frac{\hbar^2}{2 m_j} \nab^2 + \frac{\hbar^2}{2 m_j}  \frac{\nab^2\left| \psi_j\right| }{\left| \psi_{j}\right|}+ \frac{\delta E_v}{\delta n_j} \Big] \psi_j,  \\ \nonumber  
\end{align}
where $j=\{b,f\}$ and $v=\{MF,VG,LOCV\}$ denote different types of accounting for quantum corrections.

We stress that such a model for fermions is equivalent to the hydrodynamic description that is routinely utilized for interacting Fermi gases.
The underlying assumption of hydrodynamics in the considered case is the fast relaxation of the Fermi surface distortions towards a sphere centered around local hydrodynamic momentum of the collective flow.
Such a thermalization is expected in the overlapping region of the fermionic and bosonic clouds~\cite{Taylor2011}.
As a further check, thorough comparison between pseudo-Schr\"odinger and atomic-orbital approaches have been performed for various systems, including repulsive Fermi-Fermi mixtures~\cite{Grochowski2017a,Karpiuk2019} and quantum Bose-Fermi droplets~\cite{Rakshit2019,Rakshit2019a} in static and dynamic scenarios.

Nonzero temperature behavior is handled by a real time thermalization governed by nonlinear dynamics of coupled equations~\eqref{evo}.
The starting point is an excited bosonic state chosen conveniently for the following evolution and zero-temperature fermionic cloud.
This state is then evolved in a real time to the point of thermalization -- understood as a saturation of observables describing the system.
Despite a transfer of energy from bosons to fermions, during the whole evolution, both pre- and post-perturbation, fermions stay very degenerate and approximately their only excitations are the collective ones.
It has to be noted that the hydrodynamic description of fermionic collective modes has to be taken with care---quantitative predictions match the experimental values if the fermionic cloud retrieves correct, ideal gas oscillation frequency while interactions are switched off~\cite{Tao2009,Karpiuk2019}.
In the case we consider this is approximately satisfied, as the cloud exhibits frequency slightly lower than the noninteracting value, $1.83 \omega_b$ (see e.g. Eq. (23) in Ref.~\cite{Yin2006}).

\begin{figure*}[h!tbp]
	\centering
	\includegraphics[width=0.49\linewidth]{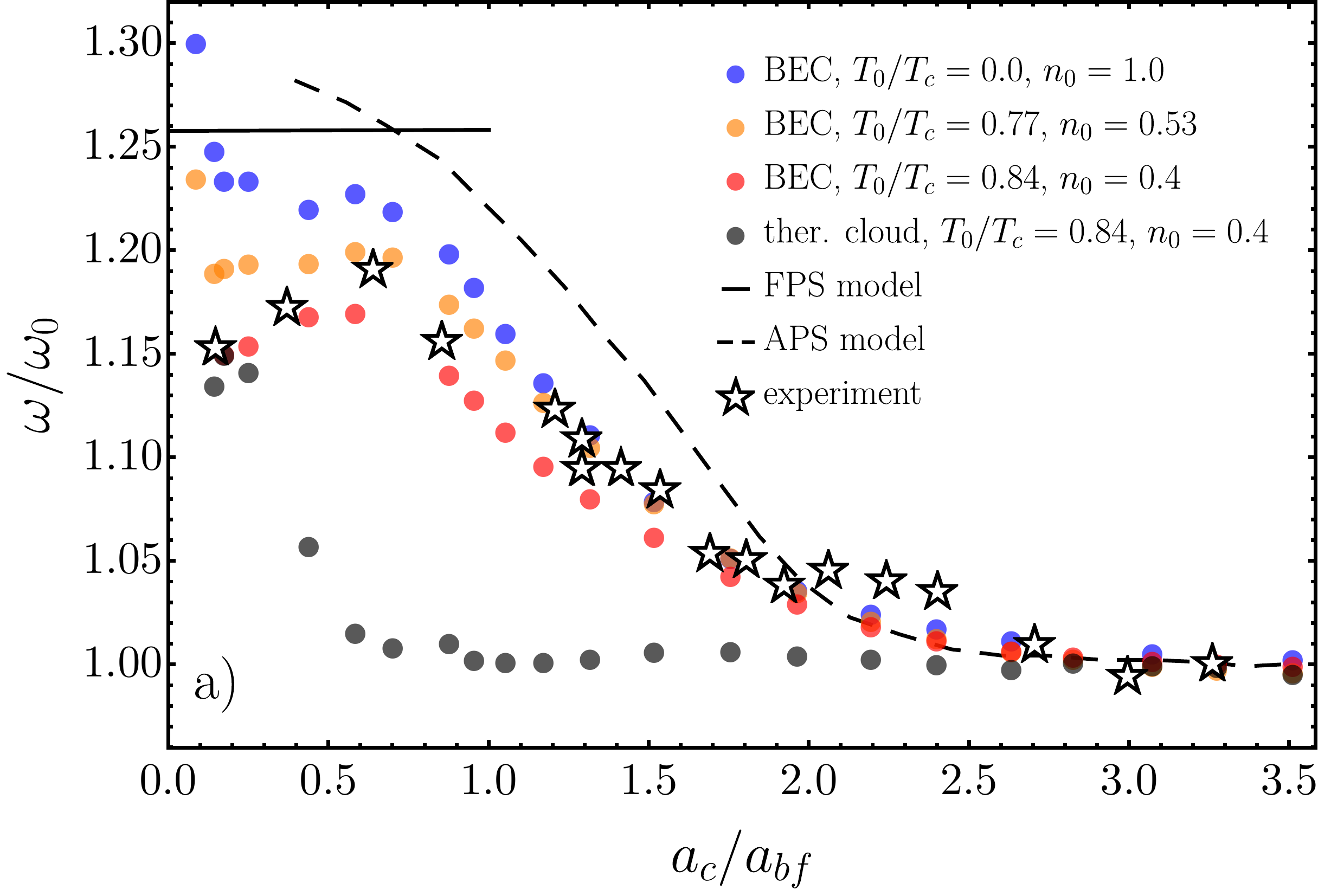}
	\includegraphics[width=0.49\linewidth]{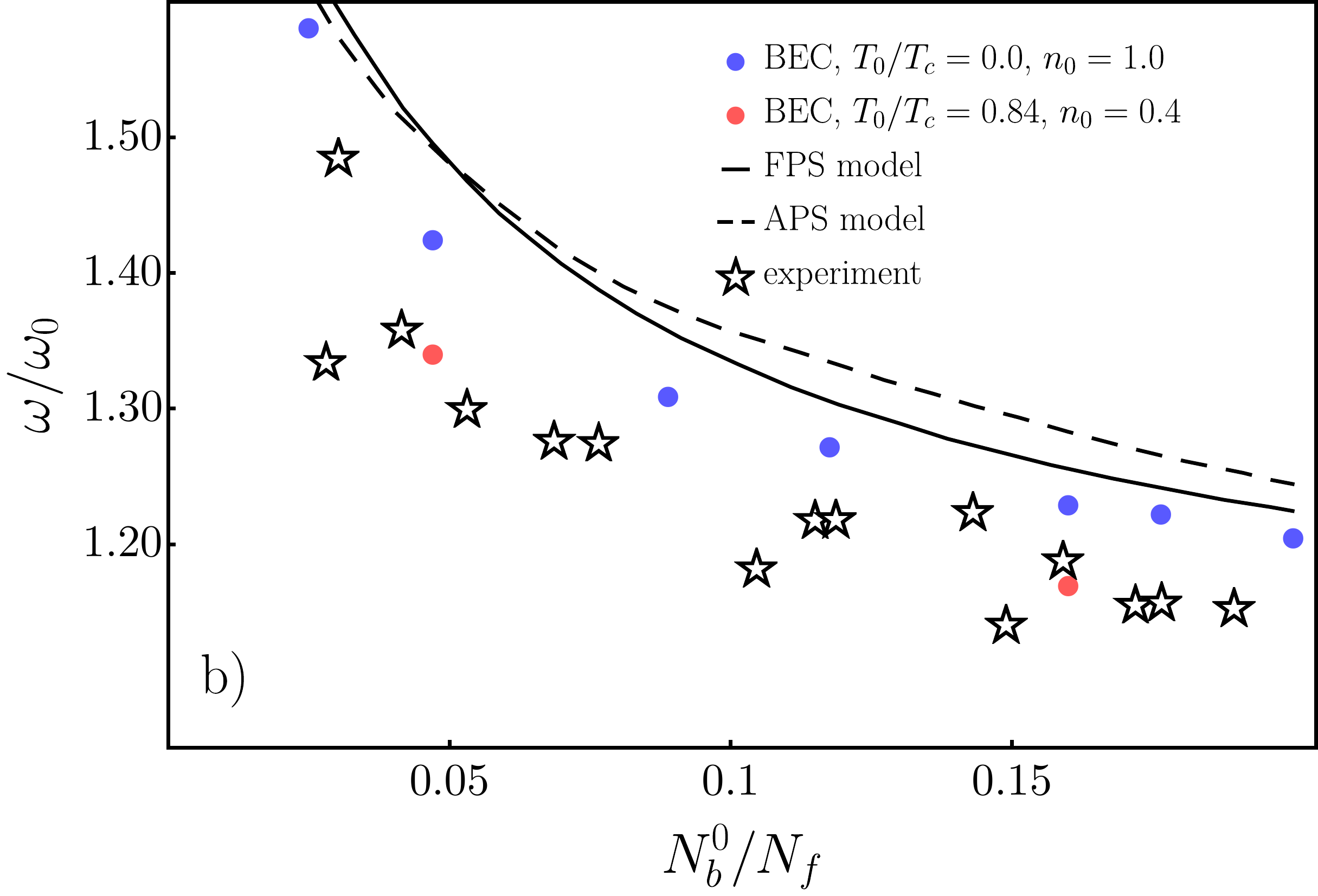}
	\caption{(a) The breathing mode frequency for a Bose-Einstein condensate immersed in a Fermi sea as a function of interaction parameter $a_c/a_{bf}$.
		The experimental data from Ref.~\cite{Huang2019} is denoted by stars.
		The theoretical data from Ref.~\cite{Huang2019} is indicated by black solid (full phase separation model) and black dashed (adiabatic Fermi sea model) lines.
		The rest of points comes from combined classical fields and pseudo-wave function model from this work.
		The blue color indicates zero temperature calculation, while the orange and red colors pertain to finite temperature ones.
		The black color signifies the oscillation frequency of the thermal cloud associated with the red markers.
		The calculations are performed within the lowest order constrained variational approach.
		For nonzero temperature calculation that involves temperature taken from the experiment~\cite{Huang2019} (red markers), the initial condensed fraction of bosons is 40$\%$ (initial temperature $T_0/T_c=0.84$) and the ratio of condensed bosons to fermions is $N_b^0/N_f=0.16$.
		The orange markers denote lower temperature, with the same $N_b^0/N_f=0.16$, but with the condensed fraction being 53$\%$ ($T_0/T_c=0.77$).
		The theory from Ref.~\cite{Huang2019} clearly overestimates the experimental curve and does not reproduce nonmonotonicity for a strong interaction.
		So does the zero temperature calculation, however to lesser extent.
		The full nonzero temperature calculation grasps the experimental data both qualitatively and quantitatively.
		For a strong interaction, up-shift of thermal cloud's frequency is observed.
		(b) The breathing mode frequency as a function of number of condensed bosons to number of fermions ratio, $N_b^0/N_{f}$.
		The interaction strength is kept constant, at $a_c/a_{bf}=0.45$.
		Analogously, the blue markers pertain to the zero temperature model and the red ones to the nonzero one.
		Again, the former overestimates the experimental data, however to the lesser extent than FPS and APS models.
		Accounting for a finite temperature allows to reproduce the experimental results.
		 \label{fig1}}
\end{figure*}

\textit{Results}---We now proceed to describe our results.
We study the breathing mode with parameters taken from the experiment and with the excitation procedure following the experimental one.
To quantify the interaction strength, critical interaction parameter is introduced, $a_c = \frac{\sqrt{15 \pi m_b m_f a_{bb}}}{2 (m_b+m_f) \sqrt{k_f^0}}$, where $k_f^0$ is the Fermi wavenumber evaluated at the center of the trap in the noninteracting case.
$a_c$ describes the value of the interaction strength, at which phase separation happens in the mean-field calculation. 
Fig.~\ref{fig1} presents the results for the number of bosons in the condensate equaling $N_b^0 = 1.6 \cdot 10^4$ and number of fermions $N_f=1 \cdot 10^5$.
The total number of bosonic atoms was chosen in such a way that the initial thermalized state has an appropriate condensed fraction for a given temperature. 
The critical parameter in this case reads $a_c = 619 a_0$ and the breathing mode frequency is normalized to $\omega_0=2 \omega_b$.

In the experiment (stars in the plot) the frequency grows from $\omega/\omega_0=1$ for the weakly interacting mixture, $a_c/a_{bf}=3.5$, starting to slightly diminish at ca. $a_c/a_{bf}=0.6$
The theory of Ref.~\cite{Huang2019} (black lines) clearly overestimates the experimental results and does not reproduce nonmonotonicity for a strong interaction.

With our framework, we find what follows:

(i) The zero-temperature version of our method lies closer to the experimental results than AFS and FPS models, however it still overestimates the frequency and does not reproduce the nonmonotonicity.
It fares better, because unlike in these two simpler approaches, it also accounts for the dynamical response of the fermionic cloud, allowing for its excitations. 

(ii) There is only very small difference between mean-field, VG and LOCV frequencies for a weak-to-moderate interaction (also within finite temperature framework).
The only appreciable (however, still within 15$\%$) mismatch is visible only very close to $a_c/a_{bf}=0$, where there is no experimental data (In Fig.~\ref{fig1} only LOCV results are presented, see the SM for MF and VG results).  
It suggests that quantum corrections play much smaller role for Bose-Fermi mixture than for Fermi-Fermi one, for which strong renormalization dependence was predicted and observed even in a weakly interacting regime.
We stress that such a lack of dependence may be specific for this particular experiment.

(iii) Introduction of a finite temperature shifts the theoretical curve much closer to the experimental one and manifests down-shift of the frequency for a very strong interaction.
For a weak-to-moderate interaction, the thermal cloud frequency stays constant at $\omega/\omega_0=1$.
As the number of noncondensed bosons is higher than condensed ones, the BEC is dragged and slowed by the thermal cloud, yielding decrease in frequency in comparison to the zero temperature case.

(iv) There is an up-shift of the thermal cloud frequency at the point for which the down-shift of the BEC frequency happens in the strongly interacting regime.

(v) The mechanism for the frequency down-shift of the BEC stems from the change of the condensed fraction of bosons.
We observe that for a strong interaction, the condensed fraction increases with time and saturates at some value (see the SM for a plot).
The stronger the interaction is, the faster it happens.
As the condensed fraction is higher, the ratio $N_b^0/N_f$ gets higher which means the decrease in a frequency (see Fig.~\ref{fig1}(b)).
The up-shift of the thermal cloud frequency can be then understood as dragging coming from, now dominating in number of atoms, condensed part of bosonic clouds that oscillate faster.

The straightforward explanation for the increase in the condensed fraction comes from the analysis of the density profiles of both species during the time evolution.
The quench into the strong interaction causes a phase separation of both clouds and effective compression of the bosonic one.
Now, due to the external pressure coming from fermions, the effective one-body trapping of bosons becomes tighter, and, as a result, their condensation temperature gets higher (as it happens for e.g. deepening of the harmonic trap in order to cool some element).

Due to the interaction quench, a rethermalization process ensues and the system starts to move toward the new equilibrium state -- the one for which the bosonic cloud is more degenerate.
To exclude the possibility of the increase of the condensed fraction being caused by additional excitations coming from the particular form of the perturbation scheme, we numerically check the adiabatic thermalization of the system.
The initial thermal state is not quenched into the strong interaction, but the change happens through a very slow interaction ramp.
Under such an evolution, for each of the final values of the interaction, the condensed fraction matches the long-time-evolved value in the quench scenario.
It supports the explanation that the frequency down-shift is caused by the rethermalization of the system.

Additionally, we have performed calculations for the other available experimental sets of data and for different values of initial condensed fractions.
The obtained results agree with the above-mentioned observations and are presented in the SM.
  
\textit{Recapitulation and outlook}---Summing up, we have created and presented a theoretical framework that allows us to effectively study quantum Bose-Fermi mixtures.
It is based on classical fields approximation from the side of the bosonic part and on the quantum continuum mechanical or hydrodynamic approach from the fermionic side.
It includes beyond mean field corrections and can be further utilized for studying physics of phase- separated states.
We have used it to describe recent experimental setting in which the breathing mode of phase-separated Bose-Fermi mixture was analyzed.
Not only did it agree with the experimental curve in the weakly coupled regime, but it also reproduced nonmonotic behavior in a strongly interacting gas, providing an insight into its underlying mechanism.
Further research in describing other quantum mixtures seems to be a natural way of continuing presented work.

\begin{acknowledgments}
\textit{Acknowledgments}---All Authors acknowledge the support from the (Polish) National Science Center Grant 2018/29/B/ST2/01308.
Part of the results were obtained using computers of the Computer Center of University of Bia\l{}ystok.
Center for Theoretical Physics of the Polish Academy of Sciences is a member of the National Laboratory of Atomic, Molecular and Optical Physics (KL FAMO).
\end{acknowledgments}

%

\end{document}